\documentclass[mathleft
]{an}
\usepackage{graphicx}
\usepackage{times}
\newcommand{\HI}{\ion{H}{I}~}
\newcommand{\HII}{\ion{H}{II}~}
\newcommand{\NH}{N$_{\rm H}$\,}
\newcommand{\NHI}{N$_{\rm HI}$\,}
\newcommand{\persec}{s$^{-1}$\,}

\newcommand{\percmc}{cm$^{-3}$\,}
\newcommand{\nms}{\mathsurround=0pt}
\newcommand{\oversim}[2]{\lower 2pt\vbox{\baselineskip 0pt \lineskip 1pt \ialign{$\nms#1\hfil##\hfil$\crcr#2\crcr\sim\crcr}}}

\newcommand{\ltsim}{\mathrel{\mathpalette\oversim{<}}}

\begin{document}

% The following seven commands are intended for editorial usage and should be ignored by
% the author(s).
\Pagespan{789}{}% Document's page range. 
% If second parameter is left empty, the last page is computed automatically.
\Yearpublication{2006}%
\Yearsubmission{2005}%
\Month{11}%   
\Volume{999}%  
\Issue{88}% 
% \DOI{This.is/not.aDOI}% 

\title{Modelling the broad-band spectra of X-ray emitting GPS galaxies}

\author{L. Ostorero\inst{1,2}\fnmsep\thanks{Corresponding author:
  \email{ostorero@ph.unito.it}\newline}
%Example 
%for footnote, note the usage of the \texttt{fnmsep}
%command as separator between institute number and footnote mark} 
\and  R. Moderski\inst{3}
\and  \L. Stawarz\inst{4,5}
\and  M.C. Begelman\inst{6}
\and  A. Diaferio\inst{1,2}
\and  I. Kowalska\inst{7}
\and  J.~Kataoka\inst{8}
\and  S.J. Wagner\inst{9}
}
\titlerunning{Modelling the broad-band spectra of X-ray emitting GPS galaxies}
\authorrunning{L. Ostorero et al.}
\institute{
       Dipartimento di Fisica Generale ``Amedeo Avogadro'', 
       Universit\`a di Torino, Via P. Giuria 1, 
       10125 Torino, Italy
\and 
       Istituto Nazionale di Fisica Nucleare (INFN), Via P. 
       Giuria 1, 10125 Torino, Italy
\and
       Nicolaus Copernicus Astronomical Center, Bartycka 18, 00-716
       Warsaw, Poland
\and 
       Kavli Institute for Particle Astrophysics and Cosmology,
       Stanford University, Stanford CA 94305
\and
       Astronomical Observatory, Jagiellonian University, ul. Orla
       171, 30-244  Krak\'ow, Poland
\and
       Joint Institute for Laboratory Astrophysics, University of Colorado, 
       Boulder, CO 80309-0440, USA
\and
       Astronomical Observatory, University of Warsaw, Al. Ujazdowskie 4, 
       00-478 Warsaw, Poland
\and
       Department of Physics, Tokyo Institute of Technology, 2-12-1,
       Ohokayama, Meguro, Tokyo, 152-8551, Japan
\and
       Landessternwarte Heidelberg-K\"onigstuhl, K\"onigstuhl 12,
       69117 Heidelberg, Germany
}

\received{}
\accepted{}
\publonline{later}


\abstract{The study of the broad-band emission of GHz-Peaked-Spectrum (GPS) 
radio galaxies is a powerful tool to investigate the physical 
processes taking place in the central, kpc-sized region of 
their active hosts, where the jets propagate and the lobes 
expand, interacting with the surrounding interstellar medium (ISM).
We recently developed a new dynamical-radiative model 
to describe the evolution of the GPS phenomenon (Stawarz et al. 2008): 
as the relativistic jets propagate through the ISM, gradually 
engulfing narrow-line emitting gas clouds along their way, the 
electron population of the expanding lobes evolves, emitting 
synchrotron light, as well as inverse-Compton radiation via 
up-scattering of the photon fields from the host galaxy and its 
active nucleus.
The model, which successfully reproduces the key features 
of the GPS radio sources as a class, provides a description of the 
evolution of their spectral energy distribution (SED) with the lobes' 
expansion, predicting significant and complex X-ray to $\gamma$-ray 
emission.
We apply here the model to the broad-band SED's of a sample of 
known, X-ray emitting GPS galaxies, and show that:
(i)  the free-free absorption 
     mechanism enables us to reproduce the 
     radio continuum at frequencies below the turnover;
(ii) the lobes' non-thermal, inverse-Compton emission can account for the 
     observed X-ray spectra, providing a viable alternative to the thermal, 
     accretion-dominated scenario.
     We also show that, in our sample, the relationship between the X-ray and 
     radio hydrogen column densitities, \NH and \NHI, is suggestive of a positive 
     correlation, which, if confirmed, would support the scenario of high-energy 
     emitting lobes.}

\maketitle

\section{Introduction}
It is currently accepted that the GPS galaxies sample the youngest fraction of the
population of powerful radio ga\-la\-xies. From sub-kpc scales, their 
jet-lobe's structures pro\-pa\-gate through the host-galaxy ISM,
evolving into sub-ga\-lac\-tic (Com\-pact Steep Spectrum, CSS) sour\-ces, 
which then expand to super-galactic scales (see O'Dea 1998 for a review).
However, this scenario still has several open issues, like    
the absorption mechanism responsible for the cha\-rac\-te\-ri\-stic turnover
in the radio spectrum, the details of the dynamical evolution and interaction 
with the ISM, the pa\-ra\-me\-ters of the central engine,
and the origin of the high-energy emission.
We recently proposed a model which addresses some of these issues 
through the analysis of the broad-band emission of GPS galaxies 
(Stawarz et al. 2008).

Here we show that our model appears to be promising, 
enabling us to reproduce a number of observed properties 
of a sample of X-ray emitting GPS galaxies.

\section{The model}

We recall below the key features of our dynamical-radiative model,
referring the reader to Stawarz et al. (2008), and re\-fe\-ren\-ces therein,
for a more comprehensive discussion.
 
Our description of the dynamical evolution of GPS sour\-ces
mounts on the model proposed by Begelman \& Cioffi (1989) to explain the 
expansion of classical double sources in an ambient medium with density 
profile $\rho(r)$. The relevant equations  can be derived by assuming that:  
(i)   the jet momentum flux 
      (proportional to the jet kinetic power $L_{\mathrm j}$) 
      is balanced by the ram pressure of the ambient medium spread 
      over an area $A_{\mathrm h}$, 
(ii)  the lobes' sideways expansion ve\-lo\-ci\-ty
      equals the speed of 
      the shock driven by the overpressured cocoon, with internal pressure $p$, 
      in the surrounding medium, and 
(iii) the energy $L_{\mathrm j} t$ transported by the jet pair during the source lifetime 
      is converted into the cocoon's internal pressure. 
For young GPS sources with age $t$, li\-near 
size  $LS(t) \ltsim 1$ kpc, and transverse 
size $l_{\mathrm c}(t)$, expan\-ding in the central 
core of the gaseous halo of the elliptical host galaxy, 
we could constrain the model with a number of reasonable assumptions:
(i)   a constant ambient density $\rho=m_p n_0$
      (with $m_p$ the proton mass, and $n_0 \simeq 0.1$ \percmc),
      representative of the King-profile's core,
(ii)  a constant hot-spot advance ve\-lo\-ci\-ty $v_{\mathrm h}\simeq 0.1c$,
      as suggested by many obervations of compact symmetric objects
      (but see Kawakatu, Nagai, \& Kino 2009 for alternative scenarios),
(iii) a scaling law $l_{\mathrm c}(t)\sim t^{1/2}$, reproducing the initial, 
      ballistic phase of the jet propagation.
All the lobes' physical quantities become thus functions of two parameters only: 
the jet kinetic power $L_{\mathrm j}$ and the source linear size $LS$.

We then studied how the 
broad-band radiative output of GPS sour\-ces evolves as the source expands,
for a given jet power $L_{\mathrm j}$.
The magnetic field in the expanding lobes scales as 
$B=(8\pi\eta_B p)^{1/2} \sim L_{\mathrm{j}}^{1/4}LS^{-1/2}$ ,
with $\eta_B=U_B/p<3$, and $U_B$ the magnetic energy density. 
The electron population $Q(\gamma)$ (with $\gamma$ the electron's Lorentz factor), 
injected from the terminal jet shocks to the ex\-pan\-ding lobes, 
evolves under the joint action of adiabatic and radiative 
energy los\-ses, yielding a lobes' electron population 
$N_e(\gamma)$, which has a broken power-law form with 
critical energy $\gamma_{\mathrm{cr}}$ when $Q(\gamma)$ is a power law, and a more
complex form when $Q(\gamma)$ is a broken power law 
with intrinsical break $\gamma_{\mathrm{int}}\simeq 2\,10^3$. 
Assuming that the lobes' electrons, in rough equipartition with the magnetic 
field and the cold protons, provide the bulk of the lobes' 
pressure, the electron energy density is $U_e=\eta_e p$, 
with $\eta_e \ltsim 3$.
The lobes'e\-le\-ctrons are source of syn\-chro\-tron radiation,
with luminosity $L_{\mathrm{syn}}$ constant with time,
and energy density $U_{\mathrm{syn}} \sim LS^{-3/2}$. 
Free-free absorption (FFA) of this radiation by 
neu\-tral-hy\-dro\-gen clouds of the narrow-line region (NLR), engulfed by the 
expanding lobes and photoionized on their surface by the radiation from the 
active nucleus (as proposed by Begelman 1999), is the process which we 
favour for the formation of the inverted spectra.
While the syn\-chro\-tron-self-absorption (SSA) process 
does not enable us to reproduce the observed turnover frequencies 
$\nu_{\mathrm{p}}$,
the spectra below the turnover, and the $\nu_{\mathrm{p}}-LS$ 
anticorrelation (O'Dea \& Baum 1997),
FFA effects best fit the inverted spectra, and are a pro\-mi\-sing 
candidate to account for the above anticorrelation.

The lobes' particles also produce inverse-Compton (IC) radiation via up-scattering
of both the synchrotron radiation (synchrotron-self-Compton mechanism; SSC) and the 
local, thermal photon fields generated by the accretion disc, the torus, and the 
stellar population of the host galaxy. 
The energy density $U_{\mathrm{rad}}$ of the thermal fields was evaluated 
by assuming that the nuclei of GPS sources share the pro\-per\-ties of quasars 
and Seyfert galaxies.
The accretion disc, assumed to produce the bulk of its luminosity
($10^{45-47}$erg \persec; e.g., Koratkar \& Blaes 1999)
at UV frequencies, provides $U_{\mathrm{UV}}\sim LS^{-2}$; the dusty torus, 
radiating the disc's UV photons at IR frequencies with efficiency 
$\epsilon\sim 10-100$\%,  yields $U_{\mathrm{IR}}\sim LS^{-2}$; finally, 
the host galaxy contributes near-IR to optical 
photons with $U_{\mathrm{opt}}$ independent of $LS$.
The IC scattering of all the above radiation fields yields significant and 
complex high-energy emission, from X-ray to $\gamma$-ray energies. 
Whereas in GPS {\it quasars} the direct X-ray emission of the accretion disc's hot
corona and of the beamed relativistic jets
may overcome the X-ray output of most of these sources, in GPS {\it galaxies} 
those contributions are expected to be obscured by the torus and
Doppler-hidden, respectively, and the lobes are expected to be the dominant X-ray source.

\section{Observational supports}
\label{sec_obssupp}
Our model, and specifically the prediction of the X-ray--emitting lobes, 
may be supported by further observational evidence, which we did not 
discuss in Stawarz et al. (2008).

The X-ray emission of GPS galaxies has been traditionally interpreted as thermal 
radiation from the accretion disc, absorbed by a gas component associated 
with the AGN and characterized by an equivalent hydrogen 
column density \NH (O'Dea et al. 2000; Guainazzi et al. 2004, 2006;
Vink et al. 2006; Siemiginowska et al. 2008),
rather than as non-thermal emission from the jets or the lobes.
The former scenario is mainly based 
on the apparent discre\-pan\-cy between the equivalent total-hydrogen column density 
\NH derived from the X-ray spectral analysis and the neutral-hydrogen column 
density \NHI derived from the 21-cm radio measurements.  
Because \NH always appeared to exceed \NHI of 1--2 orders of 
magnitudes, 
it came natural to interpret the X-rays as produced in a
source region which is more obscured than the region where the bulk of the
radio emission comes from, and thus located {\it in between} the radio lobes;
%if this were not the case, 
otherwise, an unreasonably high fraction of ionized hydrogen 
(\ion{H}{II}) would be necessary to account for the above difference 
(e.g., Guainazzi et al. 2006; Vink et al. 2006).
Such a scenario would also be consistent with the observed anticorrelation 
between \NHI and linear size found by Pihlstr\"om, Conway \& Vermeulen (2003), 
being the fraction 
of ionized gas likely low in a young radio source with still expanding 
Str\"omgren sphere (Vink et al. 2006).

The discrepancies between the \NH and \NHI values mentioned above
should actually be regarded with caution.
The \NHI estimate is derived, from the measurements of the 
\HI\linebreak absorption lines, as a function of the ratio between the 
spin temperature $T_{\mathrm{s}}$ of the gas and its covering factor $c_{\mathrm f}$, representing 
the fraction of the source covered by the \HI screen (e.g., Gupta et al. 2006).
The common assumption $T_{\mathrm{s}}/c_{\mathrm{f}} = 100$ K 
refers to the case of complete coverage ($c_{\mathrm{f}}=1$) of the emitting 
source by a standard cold ($T_{\mathrm{k}}\simeq 100$ K) ISM cloud
in thermal equilibrium, and thus with spin temperature 
equal to the kinetic temperature 
($T_{\mathrm{s}}=T_{\mathrm{k}})$.
However, this assumption returns a value of \NHI which represents 
a lower limit to the actual neutral hydrogen column density 
(Pihlstr\"om et al. 2003; Vermeulen et al. 2003; Gupta et al. 2006).
In fact, in the AGN environment, illumination by X-ray radiation might 
easily raise $T_{\mathrm{k}}$ up to $10^3-10^4$ K (Conway \& Blanco 1995; 
Maloney, Hollenbach, \& Tielens 1996), making $T_{\mathrm{s}}$ raise accordingly 
(Davies \& Cummings 1975; Liszt 2001);
a source covering factor smaller than unity would also increase 
the $T_{\mathrm{s}}/c_{\mathrm{f}}$ ratio.
Both the above effects might lead to \NHI values fully 
consistent with the \NH estimates.
Finally, temperatures as high as several $10^3$ K would likely 
imply the presence of a non-negligible fraction of \HII 
(Maloney et al. 1996; Vink et al. 2006), also contributing to
relax possible residual column-density discrepancies.
The consistency of \NH and \NHI would 
make the scenario of non-thermal X-ray--emit\-ting lo\-bes 
a viable alternative to the accretion-disc dominated mo\-del.

Evidence is mounting that, in GPS and CSS sources, the \HI absorption 
lines are not generated by a screen co\-ve\-ring the source uniformly: instead,
they seem to originate in clouds of  neutral hydrogen connected with 
the radio structures of their 
jets and/or lobes, and possibly interacting with them
(Morganti et al. 2004; Labiano et al. 2006; Vermeulen et al. 2006).
The association of the bulk of the \HI absor\-ption with the optical emission lines 
currently supports the identification of the absorbers with the atomic cores of the 
NLR clouds, although the presence of \HI elsewhere is not ruled out 
(Labiano et al. 2006; Vermeulen et al. 2006).

In our GPS model, the NLR clouds are gradually engulfed by the expanding lobes:
besides being responsible for the FFA of the radio photons, 
they might play an important role in the absorption of the lobes' X-ray radiation.

\section{Comparison with broad-band data}
\subsection{Modelling the SED's}
\label{sec_seds}
\begin{figure}
\vspace{-1.5cm}
\includegraphics[width=1.0\hsize,angle=0]{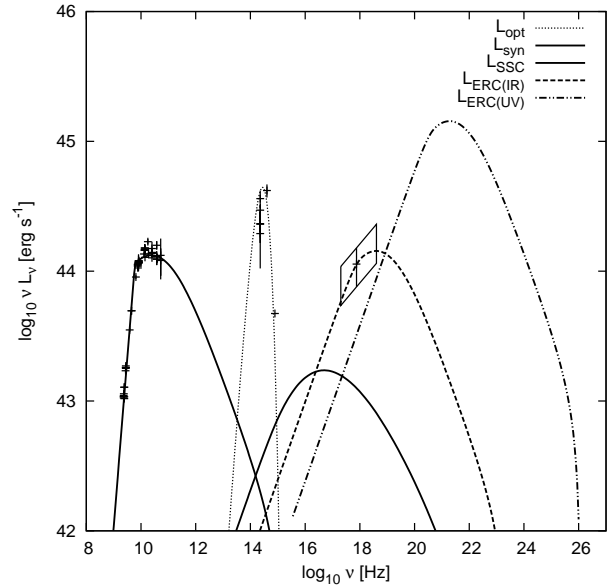}
\vspace{-2.cm}
\caption{Modelling of the intrinsic  
         SED of GPS galaxy IERS B0108+388. 
         Radio to X-ray data were derived from: NED; Dallacasa et al. (2000); 
         Tinti et al. (2005); Stickel et al. (1996); Vink et al. (2006).
         The curves show the modelled spectral components:
         synchrotron emission, 
         and corresponding SSC emission (solid lines); 
         thermal star light (dotted line); comptonized thermal 
         emission from the torus and the disc, respectively 
         (dashed and dash-dot-dotted lines). The 
         comptonized starlight's luminosity does not appear in the plot
         because it is lower than $10^{42}$ erg~\persec.}
\label{fig_seds}
\end{figure}

We tested our dynamical-radiative model on the 11 GPS galaxies currently known 
as X-ray emitters.\footnote{IERS B0026+346, IERS B0108+388$^{\star}$, 
IERS B0500+019$^{\star}$, IERS B0710+439, PKS B0941-080, IERS B1031+567$^{\star}$, 
IERS B1345+125$^{\star}$, IVS B1358+624$^{\star}$, 
IERS B1404+286, IERS B2128+048, IERS B2352+495$^{\star}$. Sources marked with an
asterisk are those of the subsample discussed in Sec.\ \ref{sec_nh}, and
included in Fig. \ref{fig_nh}.}
In Fig.\ \ref{fig_seds}, we show, as an example, the modelling of the 
intrinsic\footnote{Throughout this paper, we use the cosmological paramters:
$\Omega_\mathrm{{\Lambda}}=0.7$, $\Omega_{\mathrm{M}} = 0.3$,
with $H_{\mathrm 0} = 72$ km \persec Mpc$^{-1}$} 
broad-band SED of B0108+388,  
a source with $LS=41$ pc. 
The SED data were derived from the literature, and properly de-absorbed.
The modelling of the complete SED sample will be presented elsewhere.

The radio data of IERS B0108+388 were modelled as synchrotron radiation produced by 
a lobes' electron population $N_e(\gamma)$ derived from the evolution 
of an injected hot-spots' population $Q(\gamma)\sim \gamma^{-s}$, with $s=s_1=1.8$
for $\gamma < \gamma_{\mathrm{int}}$, and $s=s_2=3.2$ for 
$\gamma>\gamma_{\mathrm{int}}$. FFA effects enable us to best fit the spectral 
behaviour at frequencies below the $\simeq 6$ GHz turnover.
The thermal emissions from the torus (IR), the disc (UV), and the host galaxy (optical-NIR) 
were modelled as black-body spectra with the appropriate frequency peaks  
($\nu_{\mathrm{IR}}=0.5\, 10^{13}$ Hz, $\nu_{\mathrm{UV}}=2.45\,10^{15}$ Hz, 
and $\nu_{\mathrm{opt}}=2.0\,10^{14}$ Hz) and 
bolometric luminosities ($L_{\mathrm{IR}}=5.0\,10^{44}$ erg \persec, 
$L_{\mathrm{UV}}=5.0\,10^{45}$ erg \persec, and $L_{\mathrm{opt}}=6.0\,10^{44}$ erg \persec).
The comptonization of the synchrotron and thermal radiation fields yields  
the high-energy spectral components. 
For this source, 
the X-ray emission is dominated by the comptonization of the IR radiation.
Our model well reproduces the observed X-ray spectrum, and predicts significant 
$\gamma$-ray emission. 

\subsection{\NH-\NHI connection}
\label{sec_nh}
Besides the modelling of the broad-band SED's, a way 
of discriminating among different scenarios, and 
unveil the actual X-ray production site, is to compare the properties of the X-ray and
radio absorbers, i.e. the \NH and \NHI column densities.
Such a comparison can be performed either for individual sources, 
where an {\it ad hoc} increase of the $T_{\mathrm{s}}$ pa\-ra\-me\-ter can remove 
possible \NH and \NHI discrepancies,
or for a source sample, where the existence of 
a positive, significant \NH-\NHI correlation would suggest that the X-ray
and radio absorbers coincide, thus supporting the co-spatiality of the X-ray and
radio source.

\begin{figure}
\includegraphics[width=83mm,height=60mm]{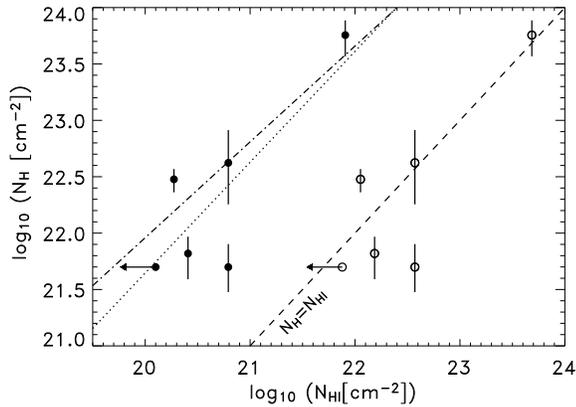}
\caption{\NH vs. \NHI for 6 GPS's of our sample$^1$.
Solid symbols: \NHI was computed with 
$T_{\mathrm{s}}=100$ K; arrows are upper limits. 
Open symbols: as an example, the same sources with
$T_{\mathrm{s}}=6\,10^3$ K are shown. 
Dash-dotted line: linear fit to the 5-source subsample 
of \NH/\NHI {\it detections} (with $T_{\mathrm{s}}=100$ K);
dotted line: linear fit to the 6-source subsample including both 
{\it detections} and {\it upper limits}; 
Data are from: 
Guainazzi et al. (2006); Mirabel (1989); O'Dea et al. (2000); Pihlstr\"om et al. (2003);
Siemiginowska et al. (2008); 
Vermeulen et al. (2003); Vink et al. (2006).}
\label{fig_nh}
\end{figure}

We investigated the existence of a connection between \NH and \NHI 
in our sample. 
For a positive correlation, we searched the source subsample
for which both \NH and \NHI\linebreak 
estimates (either detections or upper limits) 
are available 
(see Fig.\ \ref{fig_nh}, and footnote 1).
We obtained the following results:
(1) the subsample of 5 sources for which both \NH and \NHI {\it detections} are
available displays a strong (Pearson's\linebreak $r=0.997$)
and highly significant ($S=2.3 \cdot 10^{-4}$) \NH-\NHI positive correlation;
(2) in the above-mentioned subsample, the strength and significance of the 
correlation substantially decrease
when using non-parametric methods\linebreak  
(Kendall's and Spearman's correlation coefficients; e.g.,\linebreak  Press et al. 1992);
(3) the 6-source subsample including the 
\NHI {\it upper limit} also
shows, according to survival analysis techniques (ASURV, Rev.\ $1.2$; e.g., La Valley,
Isobe \& Feigelson 1992), 
a positive correlation, however with lower strength 
($\rho\sim0.6$),
and a significance varying in the range 
$S=0.17-0.33$, depending on both the
\NH value (when more than one is available) and the statistical method chosen
for the analysis.
Although the data are suggestive of a positive correlation, further measurements 
would definitely help to improve the statistics.

\section{Conclusions and future prospects}
Our dynamical-radiative model 
can reproduce the observed broad-band SED's of 
X-ray emitting GPS galaxies. 
The\linebreak shape of the radio spectra at frequencies below the 
tur\-no\-ver is best fitted by as\-su\-ming FFA effects as 
the domi\-nant absorption mechanism,
whereas the X-ray spectra can be ascribed to IC scattering 
of the thermal radiation fields (accretion disc, torus, and host galaxy) 
off the lobes' e\-lec\-tron population.
Further observational support to the X-ray--lo\-bes' scenario comes from 
the radio and X-ray hydrogen column densities of a sample of X-ray 
GPS's: the data suggest a positive correlation, which, if confirmed, 
would point towards the co-spatiality of the radio and X-ray emission sites.
Additional measurements, necessary to improve the statistics, are already 
planned.

\acknowledgements
This research has made use of the NA\-SA/I\-PAC Extragalactic Database 
(NED), which is operated by the JPL,
Caltech, under contract with the NASA.
L.O. and A.D. a\-cknow\-ledge partial support from the INFN grant PD51.
{\L}.S., R.M., J.K., and S.W. acknowledge support by MEiN grant 1-P03D-003-29, 
MNiSW grant N N203 301635, JSPS KAKENHI (19204017/\linebreak14GS0211), and BMBF/DLR 
grant 50OR0303, respectively.
This work has benefited from research funding from the European\linebreak Community's 
sixth Framework Programme under RadioNet\linebreak R113CT 2003 5058187.
We are indebted with L. Costamante, A. Sie\-mi\-gi\-no\-wska,
M. Guai\-naz\-zi, R. Morganti, and E. Ferrero for helpful discussions on GPS's. 
L.O. and \L.S. are grateful to M.J. Geller and A. Siemiginowska
for their kind hospitality at the CfA.
We thank the referees for their relevant and appropriate comments.

\end{document}